\documentclass[11pt,epsfig]{article}
\usepackage{amsfonts}
\usepackage{graphicx}
\usepackage{mathrsfs}
\usepackage{bm}
\usepackage{upgreek}
\usepackage{color}

\begin{document}

\title{Why $X(3915)$ is so narrow as a $\chi^{}_{c0}(2P)$ state}
\vspace{10mm}

\author{Yue Jiang$^{1}$\footnote{jiangure@hit.edu.cn},~~Guo-Li Wang$^{1}$\footnote{gl\_wang@hit.edu.cn},~~Tianhong Wang$^{1}$
\footnote{thwang.hit@gmail.com},~~Wan-Li Ju$^{1}$\footnote{scanh2000@126.com}
\\
\\
{\it \small $^1$ Department of Physics,
Harbin Institute of Technology, Harbin 150001, China}\\}
\date{}
\maketitle

\baselineskip=24pt

\begin{quotation}

\vspace*{1.5cm}
\begin{center}
  \begin{bf}
  ABSTRACT
  \end{bf}
\end{center}

\vspace*{0.5cm} \noindent
New resonance $X(3915)$ was identified as the charmonium $\chi_{c0}(2P)$
by BABAR Collaboration, but there seems still open question of this
assignment: why its full width is so narrow?
To answer this question, we calculate the Okubo-Zweig-Iizuka (OZI) allowed
strong decays $X(3915)\to D \bar D$, where  $X(3915)$ is assigned as a
$\chi_{c0}(2P)$ state, and estimate its full width in the cooperating
framework of $^3\!P_0^{}$ model and the Bethe-Salpeter (BS) method using
the Mandelstam Formalism, during which non-perturbative QCD effects of the
hadronic matrix elements are well considered by overlapping integral over
the relativistic Salpeter wave functions of the initial and finial states.
We find the node structure of $\chi_{c0}(2P)$ wave function resulting in
the narrow width of $X(3915)$ and show the dependence of the decay
width on the variation of the initial mass of $X(3915)$. We point out that the rate of $\frac{\Gamma(X(3915)\to D^+ D^-)}{\Gamma(X(3915)\to D^0 \bar D^0)}$ is crucial to confirm whether $X(3915)$ is the $\chi_{c0}(2P)$ state or not.

\end{quotation}

{\it Keywords:} $X(3915)$; OZI; $^3\!P_0$ Model.

PACS numbers: 12.39.Ki, 13.25.Ft

\newpage
 \setcounter{page}{1}

\section{Introduction}

The resonance $X(3915)$ was first observed by the Belle Collaboration in
two-photon collisions decaying to $J/\psi\omega$ \cite{X3915Belle}, which
indicated that it could be a conventional $c\bar c$ state. Particle data
group has listed this resonance with mass $M=(3917.5\pm2.7)\,
\mathrm{MeV}/c^2_{}$, and width $\Gamma=(27\pm10)\,\mathrm{MeV}$
\cite{pdg}.
$X(3915)$ draws a lot of attentions since it was discovered for its interesting
enigmatic features. It is first suggested by Liu {\it et al} \cite{Liu}
that it could be a $P$-wave charmonium $\chi^{}_{c0}(2P)$ state, and this
assignment has been studied extensively by Yang {\it et al} \cite{Xia};
some authors considered the possibility of it being a
$D^*_{}\bar D^*_{}$ bound state \cite{Yang, duque}; other authors point
out that it may be the same state as the $Y(3930)$\cite{yuan,nielsen}. 
Recently, BABAR Collaboration performs a spin-parity
analysis of the process  $X(3915)\to J/\psi\omega$, and they suggest that
the quantum number of this state is $J^P=0^+_{}$, therefore, the $X(3915)$
is identified as the $\chi^{}_{c0}(2P)$ state \cite{X3915Babar}.

But there are still some problems: as a $\chi^{}_{c0}(2P)$ state, the
dominant decays to $D \bar D$ of $X(3915)$ are still missing in the
$D \bar D$ invariant mass distribution; Yang {\it et al} \cite{Xia}
calculated the strong decays $X(3915)\to D \bar D$ using the simple
$^3\!P_0^{}$ model, and they find that the full width
of $\chi^{}_{c0}(2P)$ state ranges from $132$ to $187$ MeV; Guo
{\it et al} \cite{guo} also point out that the current data for the full
width of $X(3915)$ is too narrow as a $\chi^{}_{c0}(2P)$ state. These
problems show that more careful experimental researches and theoretical
studies are still needed. In this paper, we study the OZI allowed strong
decays of $X(3915)$ as the excited  $P$-wave $\chi^{}_{c0}(2P)$ state
using the combined method of $^3\!P_0^{}$ model and Bethe-Salpeter equation.

The $^3\!P_0^{}$ model is first proposed by Micu \cite{Micu}, later developed and applied
to hadron decays by Yaouanc {\it et al} \cite{Yaouanc1,Yaouanc2}. It is established on
the supposition that $q\bar q$ pair is created from the vacuum
which has the quantum number of $^{2S+1}\!L^{}_J=\,^3\!P^{}_0$.
This model has been used successfully to deal with strong decays
of light mesons \cite{Blundell, Ackleh, Barnes} and heavy mesons \cite{Liu, Zhang, Luo, Sun},
and has also been widely used for describing
OZI-allowed strong decays \cite{Okubo1, Zweig, Iizuka, Okubo2}.
Meanwhile, the Bethe-Salpeter (BS) method \cite{BS,salp} is an effective relativistic way to calculate
heavy mesons' annihilation decays \cite{Wang1, Wang2}, weak decays \cite{Fu, WangZH},
radiative decays \cite{WangTH}, during which the large relativistic
corrections for $P$-wave states are found and considered.

Combining the two powerful schemes mentioned above as in the
previous paper \cite{Fu2} of our group, we are able to deal with
the OZI-allowed strong decays in which large relativistic corrections may be involved.
In this cooperating framework of BS
method and $^3\!P_0^{}$ model, the decay
amplitude has the same structure as the $^3\!P^{}_0$
model, and the wave functions are given by the BS equations. The significance of these
decays lies on their dominance of all the $\chi^{}_{c0}(2P)$ decay modes, so we use them to estimate the full width of $\chi^{}_{c0}(2P)$.

The rest part of this paper is organized as following:
in section 2, the calculation of transition matrix at the leading order
is illustrated; Section 3 shows the BS wave functions we applied here;
we give the numerical results in Section 4, discussions and conclusions in Section 5.

\section{Transition Matrix}

The Feynman diagram of $X(3915)\rightarrow D\bar D$
is shown in Fig. 1, where $P$ and $M$ are the momentum and mass
of the initial state $X(3915)$, and so are $P^{}_D$, $P'^{}_D$, $M^{}_D$
as those of the final mesons, respectively.
The quark-antiquark pair created in the vacuum is noted as
$q$ and $\bar q$, which stand for $u\bar u$ or
$d\bar d$. In $^3P_0$ model, instead of the vacuum, the quark-antiquark pair is produced through the vertex $-ig$. $D$ means $D^+_{}$ or $D^0_{}$, and $\bar D$ stands for
$D^-_{}$ or $\bar D^0_{}$. Other quantities such as quark
masses and the corresponding momenta are all marked out around
in the figure.

\begin{figure}[hbt]\label{Feyn}
\begin{center}
\includegraphics[height=5cm]{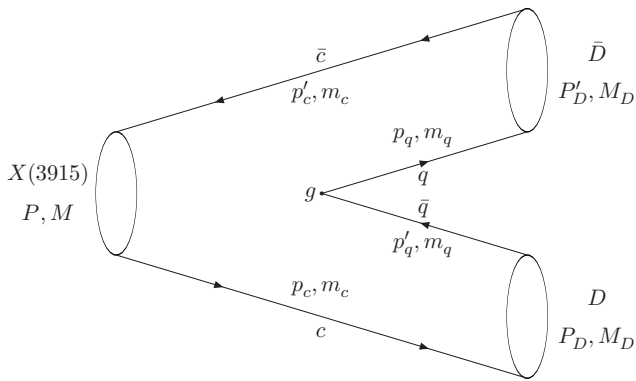}
\caption{The OZI-allowed strong decay of $X(3915)\rightarrow D\bar D$}
\end{center}
\end{figure}
According to the Mandelstam Formalism \cite{mandelstam, Kitazoe}, the
amplitude of an OZI-allowed process $X(3915)\rightarrow D\bar D$ as shown in Fig. 1 can be expressed as
\begin{equation}\label{amplitude}
{\mathcal{M}}=-ig\int\frac{\mathrm d^4_{}q^X_{}}{(2\pi)^4_{}}\mathrm{Tr}\left[\chi^X_P(q^X_{})iS^{-1}_2(p'^{}_c)\overline\chi^{\bar D}_{P'^{}_D}(q^{\bar D}_{})\overline\chi^D_{P^{}_D}(q^D_{})(-i)S^{-1}_1(p^{}_c)\right],
\end{equation}
where $\chi$ is the BS wave function, and $\overline\chi=\gamma^0_{}\chi^\dag_{}\gamma^0_{}$. For example, $\chi^X_P(q^X_{})$ stands for the BS wave function of initial meson $X$ whose momentum is $P$ and relative momentum is $q^X$, where we have the momentum relations: $p_c={\frac{m_c}{m_c+m_c}} P+q^X$, $p^{\prime}_c={\frac{m_c}{m_c+m_c}} P-q^X$. $S_i$ ($i=1,2$) is the propagator of  quark or antiquark. The
$\delta$ functions describing the momentum conservations have been
suppressed.

Since it is very difficult to solve the full BS equation to obtain the full BS wave function, and  the time component of its potential is not physically clear to us, we make an approach to Eq.~(\ref{amplitude}) and to BS equation, which is the instantaneous approximation. For the BS equation, the instantaneous approximation means that we delete the dependence of potential on the time component, and then BS equation reduces to the Salpeter equation \cite{salp}. In this case we obtain the Salpeter wave function. For Eq.~(\ref{amplitude}), we make instantaneous approach by taking the integral over $q^{X}_{_P}$, which is defined as {$q^{X}_{_P}=\frac{{q^{X}\cdot P} }{M}$}. Then we obtain the similar amplitude formula as in Ref.
\cite{Fu2} where we deal with the decays of $\Upsilon(nS)\to B \bar B~(n=4,5,6)$. If we only consider the dominant leading order contribution, the amplitude is written as
\begin{equation}\label{ampl}
{\mathcal{M}}=-ig\int\frac{\mathrm d^3_{}q^X_{_{P\perp}}}{(2\pi)^3_{}}\mathrm{Tr}\left[\frac{\not\!P}{M}\varphi^{++}_{_X}(q^X_{_{P\!\perp}})
\frac{\not\!P}{M}\overline{\varphi^{++}_{_{\bar D}}}(q^{\bar D}_{_{P\!\perp}})\overline{\varphi^{++}_{_D}}(q^D_{_{P\!\perp}})\right],
\end{equation}
where $q^X_{_{P\!\perp}}=q^{X}-q^{X}_{_{P\!\perp}}$, and because of the momentum conservation we have the following relations of the relative momenta $q^D_{_{P\!\perp}}=q^X_{_{P\!\perp}}-\alpha^D_1P^D_{_{P\!\perp}}$,
$q^{\bar D}_{_{P\!\perp}}=q^X_{_{P\!\perp}}+\alpha^{\bar D}_2P^{\bar D}_{_{P\!\perp}}$
in which $\displaystyle\alpha^D_1=\alpha^{\bar D}_2=\frac{m^{}_c}{m^{}_c+m^{}_q}$. $\varphi^{++}_{}$ or $\overline{\varphi^{++}_{}}$ is the
positive energy part of the Salpeter wave function, we will show its expression in next section.

One can find that the transition matrix element in Eq. (\ref{ampl}) is
formulated as an overlapping integral of the Salpeter wave functions of
the initial and finial states. The non-perturbative effects are included
in the wave functions since they are obtained by solving the Salpeter
equation whose kernel is a QCD-inspired potential, whose expression will be given in Section 4.

\section{Relativistic Wave Functions}

We briefly review the expressions of the Salpeter wave functions in this section. BS equation \cite{BS} or Salpeter equation \cite{salp} is a dynamical equation to relativistically deal with bound state problem and to obtain the corresponding eigenvalue and numerical value of wave function for a bound state. Though there are constrain equations on the wave function in BS method, there is no way from the BS method to give the form (or expression) of the wave function for a bound state, we have to give its expression before we put it into the equation and solve it to obtain the numerical value.

\subsection{Wave function for $D$ meson}

From the quantum field theory, the general relativistic wave function can be constructed by the meson's momentum $P$, relative momentum $q$, possible polarization vector $\epsilon^{\mu}$ or polarization tensor $\epsilon^{\mu\nu}$, and Dirac gamma matrices $\gamma$, {\it et~al}. Meson $D$ or $\bar D$ is pseudoscalar, which is also described as a $^1\!S_0$ or $0^{-}$ state.

In this way, the general expression of the Salpeter wave function which also has the quantum number of $J^{P}_{}=0^{-}_{}$ for $D$ meson (same expression for $\bar D$) in center of its
mass system can be written as \cite{cskimwang}:
\begin{equation}\label{wave}
\varphi_{_D}(\bm q')= M_{D}\left[\frac{\not\!P_{D}}{M_{D}}f^{}_1(\bm
q')+f^{}_2(\bm q')+ \frac{{\not\!q}^{\prime}\!\!_{_{P\!\perp}}}{M_D} f^{}_3(\bm
q')+\frac{\not\!P_{D}{\not\!q'}^{}\!\!_{_{P\!\perp}}}{M^2_{D}} f^{}_4(\bm
q')\right]\gamma^{}_5,
\end{equation}
where we have use $q'$ instead of $q^D$, and ${q'}^{}\!\!_{_{P\!\perp}}\!\!\!=\!(0,\,\bm q')$. The wave function $f_i(\bm q')~(i=1,...,4)$ is a function of $\bm q'^2$. One notes that there should exist other 4 terms in Eq.~(\ref{wave}) with $P_{D}\cdot {q'}^{}\!\!_{_{P\!\perp}}\!\!\!$ whose $J^P$ are also $0^-$, but they are vanished in the center of mass system because of the instantaneous approximation.

The wave function can be separated as four parts $\varphi=\varphi^{++}+\varphi^{+-}+\varphi^{-+}+\varphi^{--}$ by applying energy projection operator, where $\varphi^{++}$ is called the positive energy wave function which is dominate, $\varphi^{--}$ the negative energy wave function which can be ignored because its contribution is very small comparing with those from positive energy part, and the other two are the constraint equations mentioned above because we have $\varphi^{+-}_{}=\varphi^{-+}_{}=0$, from which we have the relations
\cite{cskimwang}:

\parbox{15.8cm}{
\begin{eqnarray*}
\hspace{1.3cm}
f^{}_3(\bm q')=-\frac{M_D(\omega^{}_{c}-\omega^{}_{q})}{m^{}_{c}\omega^{}_{q}+m^{}_{q}\omega^{}_{c}}
f^{}_2(\bm q'),
\;\;\; {f^{}_4(\bm q')}=-\frac{M_D(\omega^{}_{c}+\omega^{}_{q})}{m^{}_{c}\omega^{}_{q}+m^{}_{q}\omega^{}_{c}}
f^{}_1(\bm q'),
\end{eqnarray*}
}
\parbox{1cm}{\begin{eqnarray}\end{eqnarray}}
where $\omega_{q}=\sqrt{m^2_q +\bm q'^2}$ and $\omega_{c}=\sqrt{m^2_c +\bm q'^2}$.
The numerical values of independent wave functions $f^{}_1$ and $f^{}_2$ can be obtained by solving the Salpeter equation, and they should fulfill the normalization condition \cite{cskimwang}:
\begin{equation}\int\frac{ d{\bm q'}}{(2\pi)^3}4f^{}_1
({\bm q'})f^{}_2({\bm q'})M^2_{D}\left[
\frac{\omega^{}_{c}+\omega^{}_{q}}{m^{}_{c}+m^{}_{q}}+\frac{m^{}_{c}+m^{}_{q}}
{\omega^{}_{c}+\omega^{}_{q}}
+\frac{2 \bm
q'^2(m^{}_{c}\omega^{}_{c}+m^{}_{q}\omega^{}_{q})}{(m^{}_{c}\omega^{}_{q}+m^{}_{q}
\omega^{}_{c})^2}
\right]=2M_D.
\end{equation}

With a non-relativistic approach: $f^{}_1 = f^{}_2$ and deleting the terms proportion to $f^{}_3$ and $f^{}_4$ from which most relativistic corrections come, the relativistic wave function in Eq.~(\ref{wave}) reduce to the familiar non-relativistic one: $\varphi_{_D}(\bm q')= (M_{D}+{\not\!P_{D}})f^{}_1(\bm q')\gamma^{}_5$.

Finally, we show the expression of the positive energy progection wave function appearing in Eq.~(\ref{ampl}):
\begin{equation}
\varphi^{++}_{_D}(\bm q')=\!\frac{M_D}{2}\!\!\left[f^{}_1(\bm q')
+f^{}_2(\bm
q')\frac{m^{}_{c}+m^{}_{q}}{\omega^{}_{c}+\omega^{}_{q}}\right]\!\!\left[
\frac{\omega^{}_{c}+\omega^{}_{q}}{m^{}_{c}+m^{}_{q}}+\frac{\not\!P_D}{M_D}
-\frac{{\not\!q'}^{}\!\!_{_{P\!\perp}}
(m^{}_{c}-m^{}_{q})}{m^{}_{c}\omega^{}_{q}+m^{}_{q}\omega^{}_{c}}
+\frac{{\not\!q'}^{}\!\!_{_{P\!\perp}}{\not\!P_D}
(\omega^{}_{c}+\omega^{}_{q})}{M_D(m^{}_{c}\omega^{}_{q}+m^{}_{q}\omega^{}_{c})}\right]
\gamma^{}_5.
\end{equation}

\subsection{Wave function for $\bm {X(3915)}$ [or $\bm{\chi_{c0}(nP)}$] state}

Resonance $X(3915)$ is considered as the $\chi_{c0}(2P)$ state in this paper, which is also called $^3\!P^{}_0$ or $0^{+}_{}$ state.
The general expression of the relativistic wave function with instantaneous approach for  $X(3915)$ in its own center of
mass system can be written as \cite{pwave1}:
\begin{equation}
\varphi^{}_{_X}(\bm q)=M\left[\frac{\not\!q^{}\!_{_{P\!\perp}}}{M}g^{}_1(\bm
q)+\frac{\not\!P{\not\!q}^{}\!_{_{P\!\perp}}}{M^2_{}}g^{}_2(\bm q)+g^{}_3(\bm
q)+\frac{\not\!P}{M}g^{}_4(\bm q)\right],
\end{equation}
with the following constraint conditions of the components \cite{pwave1}:

\parbox{15cm}{
\setlength{\arraycolsep}{0.05cm}
\begin{eqnarray*}
\hspace{1.3cm}
g^{}_3(\bm q)=-\frac{{\bm q}^2}{Mm_c}g^{}_1(\bm q);
\;\;\; {g^{}_4(\bm q)}=0.
\end{eqnarray*}
}
\hfill
\parbox{1cm}{\begin{eqnarray}\end{eqnarray}}

The normalization condition for the $X(3915)$ wave function is \cite{pwave1}:
\begin{equation}
\int \frac{ d{\bm q}}{(2\pi)^3}\frac{8g^{}_1(\bm
q)g^{}_2(\bm q) \omega_c \bm
q ^2}{m_c}=2M.
\end{equation}

The expression of positive projection of the $X(3915)$ wave function is:
\begin{eqnarray}\label{0+pro}
\varphi^{++}_{_X}(\bm q)=\frac{1}{2}\left[\frac{\omega^{}_{c}}{m_c}g^{}_1({\bm q})+g^{}_2({\bm q})\right]   \left[-\frac{\bm q^2}{\omega^{}_{c}}
+\frac{m^{}_{c}}{\omega^{}_{c}}{\not\!q}^{}_{_{P\!\perp}}
+\frac{{\not\!P}}{M}{\not\!q}^{}_{_{P\!\perp}}\right].
\end{eqnarray}

\section{Numerical Result}
When we solve the Salpeter equation to obtained the eigenvalues $M$, $M_D$  and the numerical values of wave functions $f_i$, $g_i$, we choose the Cornell potential as the interaction between quark and antiquark.
In the momentum space, the potential read:

\hspace{-.7cm}\parbox{16cm}{
\setlength{\arraycolsep}{1mm}
\begin{eqnarray*}
\label{Cornell}
V(\bm q)&=&(2\pi)^3V_s(\bm q)
+\gamma_0\otimes\gamma^0 (2\pi)^3 V_v(\bm q),\\
V_{s}(\bm q)&=&-\left(\frac{\lambda}{\alpha}+V_0\right)\delta^{3}(\bm q)
+\frac{\lambda}{\pi^{2}}\frac{1}{(\bm q^{2}+\alpha^{2})^{2}},\\
V_v(\bm q)&=&-\frac{2}{3\pi^{2}}
\frac{\alpha_{s}(\bm q)}{\bm q^{2}+\alpha^{2}},\\
\alpha_s(\bm q)&=&\frac{12\pi}{27}
\frac{1}{{\rm{ln}}\left(a+\displaystyle\frac{\bm q^2}{\Lambda_{\mathrm{QCD}}}\right)},
\end{eqnarray*}
}
\hfill
\parbox{1cm}{\begin{eqnarray}\end{eqnarray}}
\newline The parameters in the potential are fixed by fitting the mass spectra of different $J^P$ state. Their values are \cite{spectrum}: $a=e=2.7183$, $\alpha$ = 0.06
GeV, $\lambda$ = 0.21 ${\rm GeV}^2$, $m_c$ = 1.62 GeV, $m_u$ = 0.305 GeV, $m_d$ = 0.311 GeV, $\Lambda_{\mathrm{QCD}}$ = 0.27 GeV.

 The value of $V_0$ is not universal for all the states. For $D$ mesons, $V_0=-0.375$ GeV. For $\chi_{c0}(nP)$ states, by fitting ground state $\chi_{c0}(1P)$ whose mass is $3414.7$ MeV, we got $V_0=-0.282$ GeV, and the mass $3837$ MeV is obtained for $\chi_{c0}(2P)$, which is about $80$ MeV lower than $X(3915)$. If we let the mass of $\chi_{c0}(2P)$ coincident with the experimental data of $X(3915)$, we get $V_0=-0.176$ GeV and $M[\chi_{c0}(2P)]=3917$ MeV. There is another parameter $g$ from $^3P_0$ model, $g=2m_q\gamma$, where $m_q$ is the mass of created quark (antiquark), $\gamma$ is a dimensionless parameter which could be served as the strength of creating a quark-antiquark pair, we choose $\gamma=0.7$ in this paper.

When the mass value $3837$ MeV of the $X(3915)$ is adopted,
we get the numerical results of the OZI-allowed strong decays as follows:

\setlength{\arraycolsep}{0.05cm}
\parbox{15.8cm}{
\begin{eqnarray*}
\Gamma[X(3837)\to D^+_{}\!D^-_{}]&=&74.9\,\mathrm{MeV},\\
\Gamma[X(3837)\to D^0_{}\bar D^0_{}]&=&59.6\,\mathrm{MeV},
\end{eqnarray*}
}
\parbox{1cm}{\begin{eqnarray}\end{eqnarray}}
and the total decay width can be estimated by the sum of these two channels:
\begin{equation}\Gamma[X(3837)\to D\bar D]=134.5\,\mathrm{MeV}.\end{equation}
When the value  $3917.5$ MeV
of the $X(3915)$'s mass is adopted, the OZI-allowed strong decays become:

\setlength{\arraycolsep}{0.05cm}
\parbox{15.8cm}{\hspace{1cm}
\begin{eqnarray*}
\Gamma[X(3915)\to D^+_{}\!D^-_{}]&=&{5.10}\,\mathrm{MeV},\\
\Gamma[X(3915)\to D^0_{}\bar D^0_{}]&=& {2.23}\,\mathrm{MeV},\\
\Gamma[X(3915)\to D\bar D]&=& {7.33}\,\mathrm{MeV}.
\end{eqnarray*}
}
\parbox{1cm}{\begin{eqnarray}\end{eqnarray}}\\
These results seem unnatural. First, usually we think the decay width to $D^+_{}D^-_{}$ and to $D^0_{}\bar D^0_{}$ of a meson should be equal, or the width of
$X(3915)\to D^+_{}D^-_{}$ is a little smaller than that of
$X(3915)\to D^0_{}\bar D^0_{}$ according to the contrast of their phase spaces,
but our results show that the decay width to $D^+_{}D^-_{}$ is obviously larger than that to $D^0_{}\bar D^0_{}$; Second, because of broader phase space, the decay width of $\chi_{c0}(2P)$ with mass $3917.5$ MeV should be larger than the case with value of $3837$ MeV, but our results are inverse, and the decay width of $\Gamma[X(3837)\to D\bar D]$ is almost 20 times larger than that of $\Gamma[X(3915)\to D\bar D]$, which is unusual. But we argue that when the wave function of initial state has a node structure, the amplitude of Eq.~(\ref{ampl}) which is overlapping integral over initial and final state wave functions could result in these unusual phenomena.

In the center of mass system of the initial state $X$, the relative momentum $q^D_{_{P\!\perp}}=q^X_{_{P\!\perp}}-\alpha^D_1P^D_{P\!\perp}$ becomes  $\bm q^D=\bm q^X-0.84\bm P^D$,
and $q^{\bar D}_{_{P\!\perp}}=q^X_{_{P\!\perp}}+\alpha^{\bar D}_2P^{\bar D}_{P\!\perp}$ becomes  $\bm q^{\bar D}=\bm q^X+0.84\bm P^{\bar D}$.
Eq.~(\ref{ampl}) becomes
\begin{equation}
{\mathcal{M}}=-ig\int\frac{\mathrm d \bm q^X}{(2\pi)^3_{}}\mathrm{Tr}\left[\frac{\not\!P}{M}\varphi^{++}_{_X}(\bm q^X)
\frac{\not\!P}{M}\overline{\varphi^{++}_{_{\bar D}}}(\bm q^X+0.84\bm P^{\bar D})\overline{\varphi^{++}_{_D}}(\bm q^X-0.84\bm P^D)\right].
\end{equation}
To show the detail of overlapping, we draw the radial wave function $f_i(\bm q)$ where $i=1,2$ for $D^+$ meson in Figure~2; since the radial wave function $g_i(\bm q)$ is always followed by $\bm q$, so we show the diagram of $\vert \bm q \vert g_i(\bm q)$ in Figure~3.

\begin{figure}[hbt]\label{0-}
\begin{center}
\includegraphics[height=7cm]{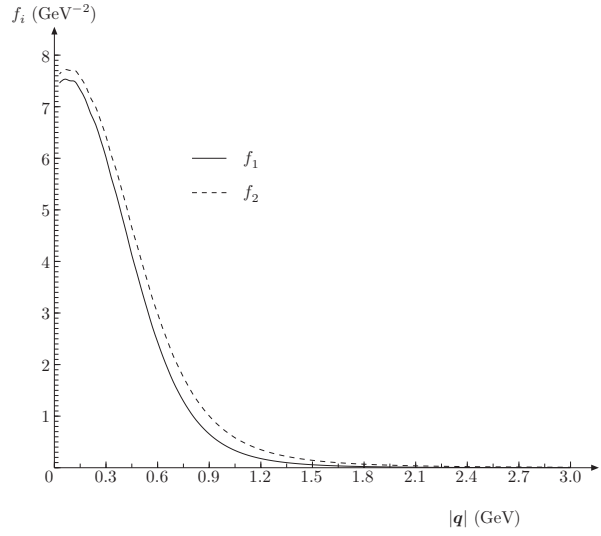}
\caption{Wave functions $f_1(\bm q)$ and $f_2(\bm q)$ of meson $D^{+}$}
\end{center}
\end{figure}

\begin{figure}[hbt]\label{0+}
\begin{center}
\includegraphics[height=8cm]{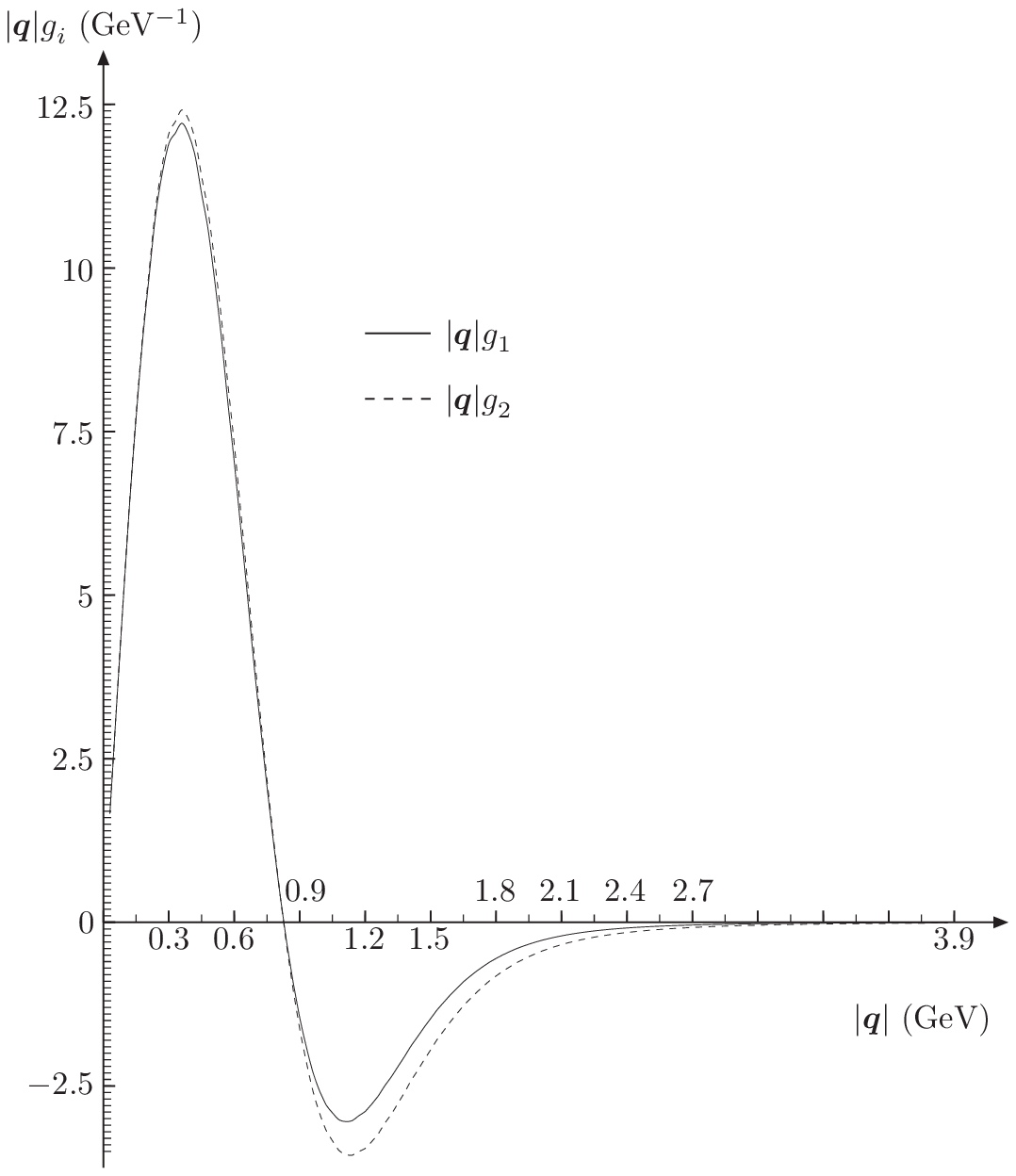}
\caption{Wave functions $\vert \bm q \vert g_1(\bm q)$ and $\vert \bm q \vert g_2(\bm q)$ of meson $X(3915)$}
\end{center}
\end{figure}

\begin{figure}[hbt]\label{Pic}
\begin{center}
\includegraphics[height=10cm]{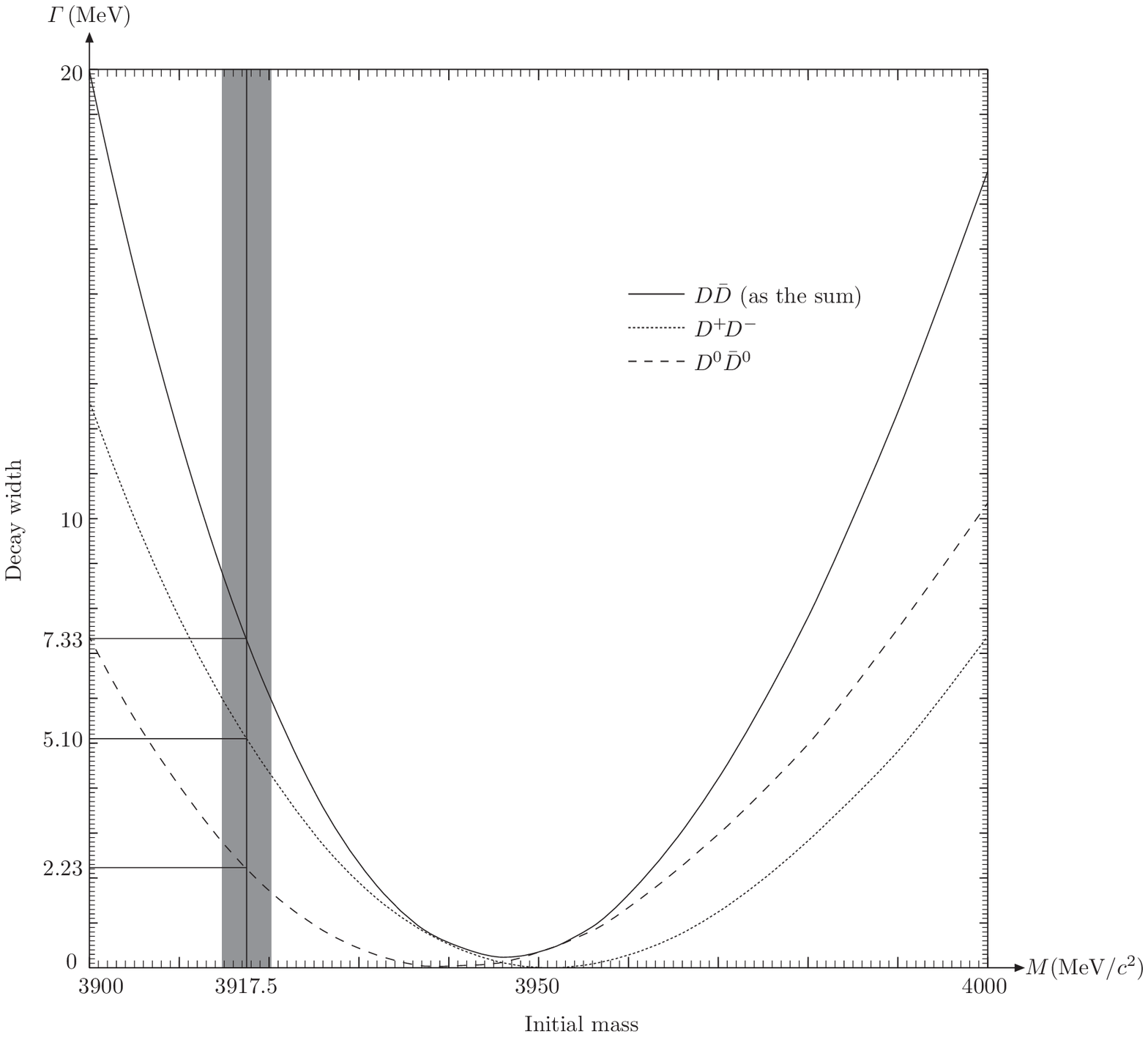}
\caption{The dependence of OZI-allowed $D\bar D$ decay width to the initial mass}
\end{center}
\end{figure}

From Figure 3, one may note that the value of $\vert \bm q \vert g_i(\bm q)$ becomes negative when the relative momentum $\bm q$ crosses the node, whose value is $810$ MeV for $X(3915)$, therefore, the negative part of wave function tends to cancel the contribution from the positive part of the same wave function. Since the amplitude is a overlapping integral of wave functions from initial and final mesons, its value or the value of the decay width is strongly affected by the momentum shifts $\bm q^X-0.84\bm P^D$ and $\bm q^X+0.84\bm P^{\bar D}$.

In the decay $X\to D \bar D$, the momentum $\bm P^D$ or $\bm P^{\bar D}$ is determined by the mass of $X$. Assuming that the $X$ mass is located at the $D\bar D$ threshold, then $\bm P^D=\bm P^{\bar D}=0$. In this case, there is no momentum shift, {\it i. e.}, the overlap of wave functions happens with no momentum shift in the full $\vert \bm q\vert $ region. From Figures~2-3, one can see that the dominant contribution comes from the small  $\vert \bm q\vert $ region because the values of $D$ meson wave function are large here, and those of the corresponding $X$ wave function with whom it overlaps are positive in this small $\vert \bm q\vert$ region,
while the contribution from negative wave function part is suppressed because of its locating at middle and large $\vert \bm q\vert $ region, and the corresponding overlaping $D$ meson wave function is small in this $\vert \bm q\vert $ region, so the overlapping integral will get its biggest value. But we point out that
 the corresponding decay width may not be large because of the limitation of phase space.

When $X$ mass becomes larger, because of the momentum shift, the overlap happens not in the full $\bm q$ region, and the contribution from negative part of $X$ wave function become larger, at the same time, the contribution from the positive part wave function becomes smaller. The overlapping integral will become smaller along with the increase of momentum shift. In Figure~4, we show the decay widths of $X\to D \bar D$ along with the changes of $X$ mass from $3900$ MeV to $4000$ MeV. One can find from the Figure 4, when $M_X\approx3940$ MeV, the decay width $X\to D^0 \bar D^0$ reaches zero,  which means the contribution of negative part wave function from $X$ meson exactly cancels the contribution from the positive part. Whereas for the decay $X\to D^+ D^-$, this happens when $M_X\approx3950$ MeV.
The experimental resonance $X(3915)$, whose mass $3917.5$ MeV is close to the cancel position, leads us to very small decay widths. When $X$ mass is $3837$ MeV, which is about $100$ MeV smaller than the cancel point mass, then we obtained very broad decay width. We notice that when $X$ mass is smaller than cancel point mass, the positive part of its wave function give dominant contribution; when its mass is larger than that mass point, the negative part will be dominant.

The node structure also can explain why we obtain different decay widths for $D^+D^-$ and $D^0\bar D^0$. The mass difference of $D^+$ and $D^0$ is about $5$ MeV, but this small difference can cause sizable difference of momentum shift in the $X$ decay. When $M_X=3917.5$ MeV, the momenta of final mesons are $\vert \bm P^{D^+}\vert=584$ MeV and $\vert \bm P^{D^0}\vert=599$ MeV respectively, which is about $15$ MeV difference. Since the $X$ mass is so close to the cancel point, the decay width is very sensitive to momentum shift and we got quite different decay widths for them: $\Gamma[X(3915)\to D^+_{}\!D^-_{}]={5.10}$ MeV, $
\Gamma[X(3915)\to D^0_{}\bar D^0_{}]= {2.23}$ MeV, {\it i. e.} the ratio \begin{equation}\frac{\Gamma[X(3915)\to D^+_{}D^-_{}]}
{\Gamma[X(3915)\to D^0_{}\bar D^0_{}]}=2.3
\end{equation} is quite large.
When $X$ mass is $3837$ MeV, which is about $100$ MeV away from the cancel point, the momenta of $\vert \bm P^{D^+}\vert=430$ MeV and $\vert \bm P^{D^0}\vert=451$ MeV with $21$ MeV difference can cause sizable different decay widths due to the node structure: $\Gamma[X(3837)\to D^+_{}\!D^-_{}]=74.9$ MeV,
$\Gamma[X(3837)\to D^0_{}\bar D^0_{}]=59.6$ MeV, but their ratio is not very large:
$$\frac{\Gamma[X(3837)\to D^+_{}D^-_{}]}
{\Gamma[X(3837)\to D^0_{}\bar D^0_{}]}=1.3
$$
So we point out that, in order to confirm that the $X(3915)$ is the $\chi_{c0}(2P)$ state, the detection of the ratio $\frac{\Gamma[X(3915)\to D^+_{}D^-_{}]}
{\Gamma[X(3915)\to D^0_{}\bar D^0_{}]}$ is a crucial one, because the node structure of $\chi_{c0}(2P)$ state will result in an unusual large ratio.

\section{Discussion and Conclusion}
Though we can explain why the resonance $X(3915)$ is so narrow as the $\chi_{c0}(2P)$ state which is suggested by BABAR Collaboration, there are still puzzles on this assignment.
We note that in Ref.~\cite{guo}, F.-K. Guo and U.-G. Meibner point out that there is an indication that the present data of the $\gamma\gamma\to D\bar D$ process already contain signals of $\chi_{c0}(2P)$ with a mass and width around $3840$ MeV and $200$ MeV. Surprisingly, this mass is exactly what we predicted in a previous paper \cite{spectrum}, in which we also correctly gave the mass of $\chi_{b0}(2P)$. And the width $200$ MeV is also fairly close to our result of $135$ MeV which we got by only
considering two channels.

In conclusion, we calculate the OZI-allowed strong decays
$X(3915)\to D^+_{}D^-_{}$ and $X(3915)\to D^0_{} \bar D^0_{}$, where $X(3915)$ is assigned as a $\chi_{c0}(2P)$ state,
in the cooperating framework of $^3\!P^{}_0$ model and the BS
method. We find the node structure in $\chi_{c0}(2P)$ wave function plays a dominant role to explain the narrow decay width of $X(3915)$. We also find big differences between $\Gamma\left[{X(3915)\to D^+_{}D^-_{}}\right]$ and $\Gamma\left[{X(3915)\to D^0_{}\bar D^0_{}}\right]$. We conclude that the ratio $\frac{\Gamma\left[X(3915)\to D^+ D^-\right]}{\Gamma\left[X(3915)\to D^0 \bar D^0\right]}$ is a crucial detection to confirm that $X(3915)$ is the $\chi_{c0}(2P)$ state in experiment.

This work was supported in part by the National Natural
Science Foundation of China (NSFC) under Grant No. 11175051.

\end{document}